\documentclass[twocolumn,prl,aps]{revtex4}
\usepackage{graphicx}
\begin{document}

\title{Quasi-one-dimensional anisotropic Heisenberg model in a
transverse magnetic field}
\author{D.V.Dmitriev}
 \email{dmitriev@deom.chph.ras.ru}
\author{V.Ya.Krivnov}
\affiliation{Joint Institute of Chemical Physics of RAS, Kosygin
str.4, 117977, Moscow, Russia.}

\begin{abstract}
The phase diagram of weakly coupled $XXZ$ chains in a transverse
magnetic field is studied using the mean-field approximation for
the interchain coupling and known exact results for an effective
one-dimensional model. Results are applied to the
quasi-one-dimensional antiferromagnet $Cs_{2}CoCl_{4}$ and the
value of interchain interaction in this compound is estimated.
\end{abstract}

\maketitle

The effects induced by magnetic fields in low-dimensional magnets
are subjects of intensive theoretical and experimental research
\cite{QPTbook}. One of the striking effects is the dependence of
magnetic properties of quasi-one-dimensional (Q1D)
antiferromagnets with anisotropic interactions on the direction of
the applied magnetic field. For example, the behavior of these
systems in a transverse magnetic field is drastically different in
comparison with the case of the longitudinal field applied along
the anisotropy axis. In particular, the transverse field induces a
gap in the spectrum and the antiferromagnetic long range order (AF
LRO) in the perpendicular direction. A quantum phase transition
takes place at some critical field, where the LRO and the gap
vanish. The phase transition of this type has been observed in the
Q1D antiferromagnet $Cs_{2}CoCl_{4}$ \cite{kenz}. The simplest
model of the one-dimensional anisotropic antiferromagnet in the
transverse field is the spin-$\frac 12$ $XXZ$ chain described by
the Hamiltonian
\begin{equation}
\mathcal{H}_{1D}=J\sum
(S_{n}^{x}S_{n+1}^{x}+S_{n}^{y}S_{n+1}^{y}+\Delta
S_{n}^{z}S_{n+1}^{z})-H\sum S_{n}^{x} \label{H1d}
\end{equation}
where $\Delta $ is an anisotropy parameter which assumed to be
$0\leq \Delta <1$.

It was proposed \cite{kenz} that low-energy properties of
$Cs_{2}CoCl_{4}$ is described by (\ref{H1d}) with $J=0.23mev$ and
$\Delta =0.25$. In contrast to the~case of the longitudinal field
the symmetry-breaking transverse field does not commute with the
$XXZ$ Hamiltonian and the exact integrability of (\ref{H1d}) is
destroyed. The model (\ref{H1d}) has been investigated using
different approximate approaches
\cite{classical,mori,hieda,Langari}. The scaling estimates at
small field \cite{XXZhx} show that the transverse field generates
the staggered magnetization $M_{st}$ $=\langle
(-1)^{n}S_{n}^{y}\rangle $ (AF LRO in the $Y$ direction) and the
gap in the spectrum $m$ (at $H=0$ the spectrum is gapless)
\begin{eqnarray}
m &\sim &(H/J)^{\frac{1}{2-d}},\qquad
d=\frac{\eta}{2}+\frac{1}{2\eta }
\nonumber \\
M_{st} &\sim &(H/J)^{\frac{\eta /2}{2-d}},\qquad
\eta=1-\frac{1}{\pi }\arccos \Delta \label{cexp}
\end{eqnarray}

To study the model (\ref{H1d}), when the field $H$ is not small,
the mean-field approximation (MFA) has been proposed in
\cite{XXZhx} and elaborated in \cite{Essler}. The MFA is based on
the \ Jordan-Wigner transformation of spin-$\frac 12$ operators to
the Fermi ones with the subsequent mean-field treatment of the
four-fermion interaction term. As a result the arising Hamiltonian
is quadratic in Fermi-operators and it is solved exactly.
Transforming this MFA\ Hamiltonian back to spin variables we
obtain a spin-$\frac 12$ $XY$ model in the longitudinal field
\begin{equation}
\mathcal{H}_{XY}=J^{\prime }\sum \left[ (1-\gamma
)S_{n}^{x}S_{n+1}^{x}+(1+\gamma
)S_{n}^{y}S_{n+1}^{y}-hS_{n}^{z}\right] \label{HXY}
\end{equation}
where parameters $J^{\prime }$, $\gamma $ and $h$ are determined
by the MFA self-consistent conditions \cite{XXZhx,Essler}.

\begin{figure}[tbp]
\includegraphics{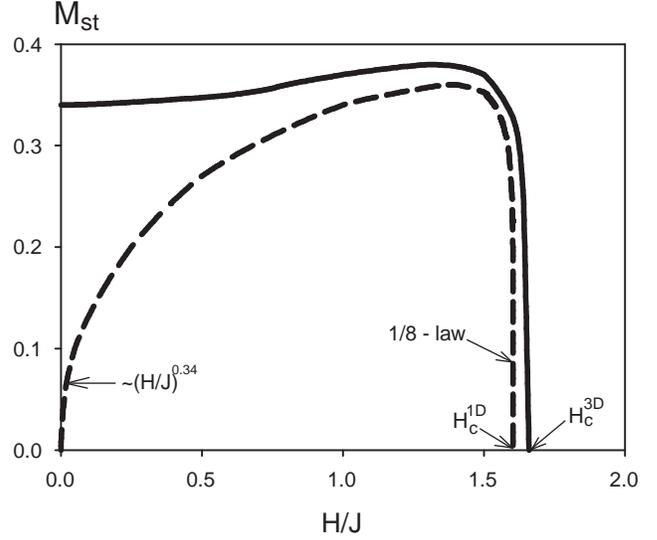}
\caption{The dependence of LRO parameter on magnetic field for
1D chain (dashed line) and quasi-1D system (solid line) for
$\Delta=0.25$.} \label{fig:My}
\end{figure}

The model (\ref{HXY}) is exactly solvable and its properties are
well studied \cite{McCoy}. This model undergoes the $T=0$ phase
transition of the 2D Ising universality class at $h=1$
corresponding to the MFA value of the critical field
$H_{c}^{1D}(\Delta )$. In particular, in the vicinity of the
critical field $M_{st}\sim \left| H_{c}^{1D}(\Delta )-H\right|
^{1/8}$. A comparison of the MFA results with those obtained in
precise numerical DMRG calculations shows high accuracy of the MFA
at $H\gtrsim J$ \cite{Essler}. The dependence $M_{st}(H)$ for
$\Delta =0.25$ obtained with use of the MFA and scaling estimate
(\ref{cexp}) is shown on Fig.\ref{fig:My} by dashed line. This
magnetization curve is qualitative similar to that observed in
neutron-scattering experiments on $Cs_{2}CoCl_{4}$. At the same
time, there is essential difference in the low-field behavior of
$M_{st}$. The experimental AF ordered moment is finite at $H=0$
while $M_{st}\to 0$ according to (\ref{cexp}). This difference is
due to weak interchain couplings in real systems and these
couplings form a 3D magnetically ordered moment below a Neel
temperature $T_{N}$. Besides, interchain couplings extend the 1D
ordered phase with $M_{st}\neq 0$ to finite temperatures.
Therefore, to describe low temperature properties of real Q1D
compounds it is necessary to take into account interchain
interactions.

In this latter we will consider the system of coupled parallel
$XXZ$ chains in the transverse field described by the Hamiltonian
\begin{eqnarray}
\mathcal{H} &=&J\sum (S_{n,\mathbf{r}}^{x}S_{n+1,\mathbf{r}
}^{x}+S_{n,r}^{y}S_{n+1,\mathbf{r}}^{y}+\Delta
S_{n,\mathbf{r}}^{z}S_{n+1,\mathbf{r}}^{z})  \nonumber \\
&&+J_{\perp }\sum
(S_{n,\mathbf{r}}^{x}S_{n,\mathbf{r}+\mathbf{\delta }
}^{x}+S_{n,\mathbf{r}}^{y}S_{n,\mathbf{r}+\mathbf{\delta}}^{y}+\Delta
S_{n,\mathbf{r}}^{z}S_{n,\mathbf{r}+\mathbf{\delta }}^{z})
\nonumber \\ &&-H\sum S_{n,\mathbf{r}}^{x}  \label{HQ1D}
\end{eqnarray}
where $n$ and $\mathbf{r}$ label lattice sites along the chain and
in perpendicular directions, $\mathbf{\delta }$ is summed over two
nearest neighbor vectors in the transverse directions, $J_{\perp
}$ is a weak coupling between neighboring chains.

A standard method for treating the model (\ref{HQ1D}) is to use
the mean-field approximation for interchain coupling and to treat
the resulting effective 1D problem as exactly as possible
\cite{Scalapino,Schulz2} (we call this approach as chain
mean-field theory (CMFT) to distinguish it from the MFA for the 1D
model (\ref{H1d})). We assume that AF order in each chain to be
oriented along the $Y$ direction and the uniform magnetization
along the $X$ axis as it occurs in the pure 1D model (\ref{H1d}).
The quasi-1D model contains another mechanism to generate the LRO.
If one of the chains is AF ordered, the interchain couplings
induce an effective staggered field on the nearest chains. In the
CMFT interchain coupling is replaced by effective fields and the
Hamiltonian (\ref{HQ1D}) reduces to an effective 1D Hamiltonian
having the form
\begin{eqnarray}
\mathcal{H}_{\mathrm{eff}} &=&J\sum
(S_{n}^{x}S_{n+1}^{x}+S_{n}^{y}S_{n+1}^{y}+\Delta
S_{n}^{z}S_{n+1}^{z})
\nonumber \\
&&-\left( H-H_{x}\right) \sum S_{n}^{x}-H_{y}\sum
(-1)^{n}S_{n}^{y} \label{Heff}
\end{eqnarray}
where fields $H_{x}$ and $H_{y}$ are determined by
self-consistency relations
\begin{equation}
H_{x}=zJ_{\perp }\left\langle S_{n}^{x}\right\rangle ,\quad
H_{y}=zJ_{\perp }M_{st}  \label{self}
\end{equation}
$z$ is the transverse coordination number.

At first, we consider the model (\ref{Heff}) at $H=0$ and $T=0$.
It can be easily shown that the self-consistency relation gives
$\left\langle S_{n}^{x}\right\rangle =0$ and the model
(\ref{Heff}) reduces to the $XXZ$ chain in the staggered field.
The low-energy properties of this model are described by a quantum
sine-Gordon model \cite{Affleck}
\begin{eqnarray}
\mathcal{H} &=&\mathcal{H}_{0}+V  \nonumber \\
\mathcal{H}_{0} &=&\frac{v(\eta )}{2}\int dx\{(\partial _{x}\Phi
)^{2}+(\partial _{x}\Theta )^{2}\}\quad   \nonumber \\
V &=&-H_{y}\sqrt{2A(\eta )}\int dx\sin (\sqrt{2\pi \eta }\Theta )
\label{Hboson}
\end{eqnarray}
where $\Phi (x)$ and $\Theta (x)$ are boson and dual fields
respectively, $v(\eta )=\frac{J\sin (\pi \eta )}{2-2\eta }$ is the
sound velocity and the coefficient $A(\eta )$ was found in
\cite{Lukyanov}.
\begin{eqnarray}
&&A=\frac{1}{8(1-\eta )^{2}}\left[ \frac{\Gamma \left( \frac{\eta }{%
2(1-\eta )}\right) }{2\sqrt{\pi }\Gamma \left( \frac{1}{2(1-\eta )}\right) }%
\right] ^{\eta }\times   \nonumber \\
&&\exp \left[ -\int_{0}^{\infty }\frac{dt}{t}\left( \frac{\sinh (\eta t)}{%
\sinh (t)\cosh \left[ (1-\eta )t\right] }-\eta e^{-2t}\right)
\right] \label{Ax}
\end{eqnarray}

The spectrum of $\mathcal{H}_{0}$ is gapless. The perturbation $V$
has scaling dimension $\eta /2$ and generates the mass gap
\begin{equation}
m=v\left( \frac{CH_{y}}{v}\right) ^{\frac{1}{2-\eta /2}}
\label{m}
\end{equation}
where constant $C$ is \cite{zam}
\begin{equation}
C=\frac{\sqrt{2A}\pi \Gamma \left( 1-\frac{\eta }{4}\right)
}{2\Gamma \left( \frac{\eta }{4}\right) }\left( \frac{2}{\sqrt{\pi
}}\frac{\Gamma \left( \frac{\eta }{8-2\eta }\right) }{\Gamma
\left( \frac{2}{4-\eta } \right) }\right) ^{2-\eta /2}  \label{C}
\end{equation}

The staggered magnetization $M_{st}$ is related to a mass gap $m$
as \cite{Lukyanov}
\begin{equation}
M_{st}=\sqrt{2A}\left\langle \exp \left( i\sqrt{2\pi \eta }\Theta
\right) \right\rangle =D\left( \frac{m}{v}\right) ^{\eta /2}
\label{sigmay}
\end{equation}
where
\begin{eqnarray}
D =\frac{\frac{\pi\sqrt{2A}}{16-4\eta} \Gamma \left( 1-\frac{\eta
}{4}\right) }{\sin \left( \frac{\pi \eta }{4-\eta }\right) \Gamma
\left( \frac{\eta }{4}\right) } \left( \frac{\Gamma \left(
\frac{2}{4-\eta }\right) \Gamma \left( \frac{8-3\eta }{8-2\eta
}\right) }{4\sqrt{\pi }}\right) ^{\frac{\eta }{2}-2} \label{D}
\end{eqnarray}

From the equations (\ref{m}) and (\ref{sigmay}) we get
\begin{eqnarray}
M_{st} &=&D\left( CD\frac{zJ_{\perp }}{v}\right) ^{\frac{\eta
/2}{2-\eta }} \label{MstH0} \\
m &=&v\left( CD\frac{zJ_{\perp }}{v}\right) ^{\frac{1}{2-\eta }}
\label{mH0}
\end{eqnarray}

The AF LRO $M_{st}$ survives at $T<T_{N}$. The Neel temperature
$T_{N}$ can be found using the random phase approximation (RPA).
The RPA dynamical susceptibility of coupled chains in disordered
phase ($T>T_{N}$) is
\begin{equation}
\chi ^{yy}(\omega ,k,k_{\perp })=\frac{\chi _{1D}^{yy}(\omega
,k)}{1-J_{\perp }(k_{\perp })\chi _{1D}^{yy}(\omega ,k)}
\label{chiRPA}
\end{equation}

The condition determined $T_{N}$ is
\begin{equation}
zJ_{\perp }\chi _{1D}^{yy}(0,\pi )=1  \label{TcRPA}
\end{equation}

The dynamical susceptibility of the 1D $XXZ$ model at $T\ll J$ is
known \cite {Schulz}
\begin{equation}
\chi _{1D}^{yy}(0,\pi )=\frac{B}{v}\left( \frac{v}{2\pi T}\right)
^{2-\eta } \label{chi0Pi}
\end{equation}
where
\begin{equation}
B=A\sin \left( \pi \eta /2\right) \frac{\Gamma ^{2}\left( 1-\eta
/2\right) \Gamma ^{2}\left( \eta /4\right) }{\Gamma ^{2}\left(
1-\eta /4\right) }  \label{B}
\end{equation}

Using the condition (\ref{TcRPA}) we extract the Neel temperature
at $H=0$
\begin{equation}
T_{N}(H=0)=\frac{v}{2\pi }\left( \frac{BzJ_{\perp }}{v}\right)
^{\frac{1}{2-\eta }}  \label{TcH0}
\end{equation}

We note that the ratio $T_{N}/m$ does not depend on $J_{\perp }$
and is determined by 1D parameter $\eta $ only.

An analysis of experimental data carried out in Ref.\cite{kenz}
has shown that the quasi-1D antiferromagnet $Cs_{2}CoCl_{4}$
consists of two interpenetrating sublattices with identical
intrasublattice interactions. These sublattices are
non-interacting on the CMFT level. Each sublattice has tetragonal
symmetry and described by the model (\ref{HQ1D}) with $z=4$.
However, no direct experimental data on the value of the
interchain interaction $J_{\perp }$ is available. The Neel
temperature in $Cs_{2}CoCl_{4}$ at $H=0$ is $T_{N}=0.0813J=0.217K$
\cite{kenz}.\ Using these data we can estimate unknown value of
$J_{\perp }$ in $Cs_{2}CoCl_{4}$. Substituting $\Delta =0.25$ ($A
=0.1405$) in (\ref{TcH0}) we find
\begin{equation}
\frac{J_{\perp }}{J}=0.0147  \label{Jp}
\end{equation}

This value is really small, so that our assumption about quasi-1D
behavior of the system is justified.

Further, using the found value of $J_{\perp }$ we can find the
staggered magnetization $M_{st}$ at $T=0$. According to
Eq.(\ref{MstH0}) $M_{st}=0.348$. The experimental value of the AF
ordered moment at $T\ll T_{N}$ is $M_{st}\approx 0.342$
\cite{kenz}. Such a perfect coincidence confirms our estimate
(\ref{Jp}). Besides, the found value of $J_{\perp }$ gives us also
the gap (\ref{mH0}) $m=0.78K$. It is remarkable that even so small
interchain coupling as in Eq.(\ref{Jp}) causes so large value of
LRO.

At $H=0$ and $T=0$ the AF LRO is generated by the interchain
couplings. At $H>0$ the 'one-dimensional' mechanism is switched.
The crude estimation of the value $H^{\ast }$, at which this
mechanism becomes predominant, can be obtained by a comparison of
(\ref{cexp}) with (\ref{mH0})
\begin{equation}
H^{\ast }\sim J\left( zJ_{\perp }/J\right) ^{\frac{2-d}{2-\eta }}
\label{hstar}
\end{equation}

At $H>H^{\ast }$ in the Hamiltonian (\ref{Heff}) the mean field
$H_{x}$ can be neglected in comparison with $H$ and at $T=0$ the
main effect of $H_{y}$ consists in a small shift of the critical
field $H_{c}^{1D}$ (see below).

At $H=H_{c}^{1D}$ and $H_y=0$ the spectrum of the model
(\ref{Heff}) is gapless. The perturbation $H_{y}\sum
(-1)^{n}S_{n}^{y}$ has scaling dimension $1/8$ and generates the
mass gap $m\sim (H_{y}/J)^{8/15}$ and AF LRO $M_{st}\sim
(H_{y}/J)^{1/15}$ in the model (\ref{Heff}). The self-consistency
relations (\ref{self}) therefore give
\begin{eqnarray}
M_{st}(H_{c}^{1D}) &\sim &\left( zJ_{\perp }/J\right) ^{1/14}
\nonumber
\\
m(H_{c}^{1D}) &\sim & \left( zJ_{\perp }/J\right) ^{4/7}
\label{sigmamhc}
\end{eqnarray}

To estimate the Neel temperature $T_{N}(H)$ in the RPA it is
necessary to know the finite temperature staggered susceptibility
$\chi _{1D}^{yy}(0,\pi ) $ for the model (\ref{H1d}) at $H>0$.
Unfortunately, it is unknown. Instead, we consider the MFA model
(\ref{HXY}), for which the susceptibility can be found. As it was
noted above the MFA describes correctly the ground state
properties of the model (\ref{H1d}) at $H\geq J$. We expect that
the MFA gives a satisfactory description of (\ref{H1d}) at low
temperature ($T\ll J)$ as well. The problem of finding $T_{N}(H)$
can be solved in the same manner as it was done by Carr and
Tsvelik in \cite{Carr} for quasi -1D quantum Ising model, which on
the CMFT\ level reduces to (\ref{HXY}) with $\gamma =1$.

We are mainly interested in the region of the fields near the
critical field $H_{c}^{1D}(\Delta )$ or at $h\sim 1$ in terms of
the MFA Hamiltonian (\ref {HXY}). Exactly at $h=1$, where the
model (\ref{HXY}) is critical, the staggered susceptibility at
$T\ll J$ according to \cite{Schulz} is
\begin{equation}
\chi _{1D}^{yy}(0,\pi )=R(\gamma )\frac{\pi }{v_{c}}\left(
\frac{v_{c}}{2\pi T}\right) ^{7/4}\frac{\Gamma \left( 7/8\right)
\Gamma ^{2}\left( 1/16\right) }{\Gamma \left( 1/8\right) \Gamma
^{2}\left( 15/16\right) }  \label{chiT}
\end{equation}
where sound velocity at the critical field $v_{c}=\gamma J^{\prime
}$ is determined from the MFA self-consistent equations
\cite{XXZhx} and
\begin{equation}
R(\gamma )=\frac{e^{1/4}2^{1/12}A^{-3}}{4}\frac{2\gamma
^{3/4}}{1+\gamma } \label{R}
\end{equation}
with Glaisher constant $A\simeq 1.282$.

The Neel temperature in the RPA is
\begin{equation}
T_{N}(H_{c}^{1D})=\frac{v_{c}}{2\pi }\left[ \frac{\pi zJ_{\perp
}}{v_{c}} \frac{R(\gamma )\Gamma \left( 7/8\right) \Gamma
^{2}\left( 1/16\right) }{ \Gamma \left( 1/8\right) \Gamma
^{2}\left( 15/16\right) }\right] ^{4/7} \label{TcHc}
\end{equation}

For $\Delta =0.25$ the critical field in the MFA is
$H_{c}^{1D}\approx 1.6J$ and $v_{c}\approx 0.185J$. Therefore, the
estimated Neel temperature for $Cs_{2}CoCl_{4}$ is
$T_{N}(H_{c}^{1D})=0.145K$.

Near the critical field at $H\gtrsim H_{c}^{1D}$ (disorder region
in 1D model) the low-temperature staggered susceptibility is well
approximated by the formula \cite{QPTbook}
\begin{equation}
\chi _{1D}^{yy}(\omega ,\pi -k)\approx \frac{\gamma
^{3/4}}{1+\gamma }\frac{v(2m/v)^{1/4}}{v_{c}^{2}k^{2}+m^{2}-\left(
\omega +i/\tau _{c}\right) ^{2}} \label{chi-tau}
\end{equation}
where the gap $m=\left| H-H_{c}^{1D}\right| $ and the phase
relaxation time $\tau _{c}=\frac{\pi }{2T}e^{m/T}$. In this case
the RPA condition of the phase transition (\ref{TcRPA}) reads
\begin{equation}
m^{2}+\tau _{c}^{-2}=\frac{\left( \gamma v_{c}\right)
^{3/4}}{1+\gamma } \left( 2m\right) ^{1/4}zJ_{\perp }
\label{RPAeq}
\end{equation}

\begin{figure}[tbp]
\includegraphics{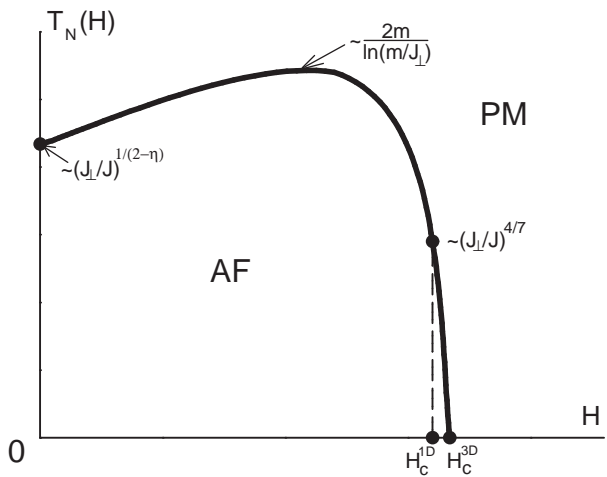}
\caption{Expected phase diagram of the model (\ref{HQ1D}).}
\label{fig:Tc}
\end{figure}

At first we estimate the shift of the critical field $\delta
H_{c}=H_{c}^{3D}-H_{c}^{1D}$ caused by interchain couplings. This
shift is determined by the condition $T\to 0$ in Eq.(\ref{RPAeq})
\begin{equation}
\delta H_{c}=2^{1/7}\frac{\left( \gamma v_{c}\right)
^{3/7}}{\left( 1+\gamma \right) ^{4/7}}\left( zJ_{\perp }\right)
^{4/7}  \label{dh}
\end{equation}

For $\Delta =0.25$ and found value of $J_{\perp }$ (\ref{Jp}) the
shift of the critical field is about $3\%$. We note that in the
vicinity of the critical point $H_{c}^{3D}$ the low-energy
properties of Q1D model (\ref{HQ1D}) belong to the universality
class of the (3+1)-dimensional classical Ising model.

Eq.(\ref{RPAeq}) gives also the behavior of Neel temperature near
the 3D critical point $H_{c}^{3D}-H\ll \delta H_{c}$
\begin{equation}
T_{N}(H)\approx 2\delta H_{c}\ln^{-1} \left( \frac{\delta
H_{c}}{H_{c}^{3D}-H}\right)   \label{Tc(h)}
\end{equation}

At intermediate fields $H^{\ast }<H<H_{c}^{1D}$ (1D ordered
region) the low temperature $T\ll m$ staggered susceptibility has
an exponential form \cite{QPTbook}
\begin{equation}
\chi _{1D}^{yy}(0,\pi )\sim (m/v)^{1/4}\xi _{c}\tau _{c}\sim
\frac{v^{3/4}}{m^{1/4}T^{3/2}}e^{2m/T}  \label{chiexp}
\end{equation}
with correlation length $\xi _{c}=v\sqrt{\frac{\pi }{2mT}}e^{m/T}$
\cite{QPTbook}. Thus, for $zJ_{\perp }\ll m(m/v)^{3/4}$ the RPA
criteria (\ref{TcRPA}) yields
\begin{equation}
T_{N}(H)\sim 2m(H)\ln ^{-1}\left( \frac{m(H)}{zJ_{\perp }}\right)
\label{Tclog}
\end{equation}

Combining the found expressions for Neel temperature in different
regions (\ref{TcH0}), (\ref{TcHc}), (\ref{Tc(h)}), (\ref{Tclog})
we arrive at the phase diagram schematically shown on
Fig.\ref{fig:Tc}. Since the gap $m(H)$ in the AF ordered region
has a maximum at some intermediate value of field
\cite{XXZhx,Essler}, then according to Eq.(\ref{Tclog}) the
function $T_{N}(H)$ also has a maximum as shown on
Fig.\ref{fig:Tc}. This fact was experimentally observed in
$Cs_{2}CoCl_{4}$ \cite{kenz}.

In summary, the main effect of interchain coupling on the system
of coupled $XXZ$ chains consists in extending of the AF ordered
phase to finite temperatures. Based on known exact results for the
effective Q1D model we estimated the value of interchain coupling
in $Cs_{2}CoCl_{4}$, which turns out to be very small. We found
that even weak coupling between $XXZ$ chains can cause a
remarkably large AF LRO below the Neel temperature.

In the presence of transverse magnetic field the 3D effects become
less pronounced. They lead to a small shift of the critical field
and to a small correction to AF LRO. The found small interchain
coupling in $Cs_{2}CoCl_{4}$ implies that the properties of this
compound in transverse magnetic field are similar to that of 1D
quantum Ising model in a magnetic field. The spectrum of this
model at exactly solvable critical point consists of eight
particles which may be observed as a few coherent peaks in the
dynamical magnetic susceptibility of $Cs_{2}CoCl_{4}$ in the AF
region slightly below the observed critical field \cite{Carr}.

This work was supported under RFBR Grant No 03-03-32141.


\begin{thebibliography}{99}
\bibitem{QPTbook}  S.Sachdev, \textit{Quantum Phase Transitions} (Cambridge
University Press, Cambridge, UK, 1999).

\bibitem{kenz}  M.Kenzelmann, R.Coldea, D.A.Tennant, D.Visser, M.Hofmann,
P.Smeibidl and Z.Tylczynski, Phys. Rev. B 65, 144432 (2002).

\bibitem{classical}  J.Kurmann, H.Tomas and G.Muller, Physica A112, 235
(1982); G.Muller and R.E.Shrock, Phys. Rev. B32, 5845 (1985).

\bibitem{mori}  S.Mori, J.-J.Kim and I.Harada, J. Phys. Soc. Jpn 64, 3409
(1995).

\bibitem{hieda}  Y.Hieida, K.Okunishi and Y.Akutsu, Phys. Rev. B64, 224422
(2001).

\bibitem{Langari} A.Langari, Phys. Rev. B69, 100402 (2004).

\bibitem{XXZhx}  D.V.Dmitriev, V.Ya.Krivnov, A.A.Ovchinnikov and
A.Langari, JETP 95, 538 (2002); D.V.Dmitriev, V.Ya.Krivnov and
A.A.Ovchinnikov, Phys. Rev. B65, 172409 (2002).

\bibitem{Essler}  J.-S.Caux, F.H.L.Essler and U.Low, Phys. Rev. B68, 134431
(2003).

\bibitem{McCoy}  E.Barouch and B.M.McCoy, Phys. Rev. A3, 786 (1971).

\bibitem{Scalapino}  D.J.Scalapino, Y.Imry and P.Pincus, Phys. Rev. B11, 2042
(1975).

\bibitem{Schulz2}  H.J.Schulz, Phys. Rev. Lett. 77, 2790 (1996).

\bibitem{Affleck}  I.Affleck and M.Oshikawa, Phys. Rev. B60, 1038 (1999).

\bibitem{Lukyanov}  S.Lukyanov and A.Zamolodchikov, Nucl. Phys. B493, 571
(1997).

\bibitem{zam}  Al.B.Zamolodchikov, Int. J. Mod. Phys A10, 1125 (1995).

\bibitem{Schulz}  H.J.Schulz and C.Bourbonnais, Phys. Rev. B27, 5856 (1983).

\bibitem{Carr}  S.T.Carr and A.M.Tsvelik, Phys. Rev. Lett. 90, 177206 (2003).

\end{thebibliography}
\end{document}